\documentclass[twoside,twocolumn,9pt]{article}
\usepackage{extsizes}
\usepackage[super,sort&compress,comma]{natbib} 
\usepackage[version=3]{mhchem}
\usepackage[left=1.5cm, right=1.5cm, top=1.785cm, bottom=2.0cm]{geometry}
\usepackage{balance}
\usepackage{mathptmx}
\usepackage{sectsty}
\usepackage{graphicx} 
\usepackage{lastpage}
\usepackage[format=plain,justification=justified,singlelinecheck=false,font={stretch=1.125,small,sf},labelfont=bf,labelsep=space]{caption}
\usepackage{float}
\usepackage{fancyhdr}
\usepackage{fnpos}
\usepackage[english]{babel}
\addto{\captionsenglish}{%

}
\usepackage{array}
\usepackage{droidsans}
\usepackage{charter}
\usepackage[T1]{fontenc}
\usepackage[usenames,dvipsnames]{xcolor}
\usepackage{setspace}
\usepackage[compact]{titlesec}
\usepackage{epstopdf}%This line makes .eps figures into .pdf - please comment out if not required.
\definecolor{cream}{RGB}{222,217,201}

\usepackage{physics} %allow braket notation and more
\usepackage{hyphenat} % allows to break compound words between lines use: \hyp{}
\usepackage[bookmarks=false,bookmarksopen=false,pdfpagelayout=SinglePage,pdfstartview=FitH]{hyperref}
\usepackage[dvipsnames]{xcolor}
\usepackage{url}
\begin{document}
%TC:break Preamble
\pagestyle{fancy}
\thispagestyle{plain}
\fancypagestyle{plain}{
%%%HEADER%%%
\renewcommand{\headrulewidth}{0pt}
}
%%%END OF HEADER%%%

%%%PAGE SETUP - Please do not change any commands within this section%%%
\makeFNbottom
\makeatletter
\renewcommand\LARGE{\@setfontsize\LARGE{15pt}{17}}
\renewcommand\Large{\@setfontsize\Large{12pt}{14}}
\renewcommand\large{\@setfontsize\large{10pt}{12}}
\renewcommand\footnotesize{\@setfontsize\footnotesize{7pt}{10}}
\makeatother

\renewcommand{\thefootnote}{\fnsymbol{footnote}}
\renewcommand\footnoterule{\vspace*{1pt}% 
\color{cream}\hrule width 3.5in height 0.4pt \color{black}\vspace*{5pt}} 
\setcounter{secnumdepth}{5}

\makeatletter 
\renewcommand\@biblabel[1]{#1}            
\renewcommand\@makefntext[1]% 
{\noindent\makebox[0pt][r]{\@thefnmark\,}#1}
\makeatother 
\renewcommand{\figurename}{\small{Fig.}~}
\sectionfont{\sffamily\Large}
\subsectionfont{\normalsize}
\subsubsectionfont{\bf}
\setstretch{1.125} %In particular, please do not alter this line.
\setlength{\skip\footins}{0.8cm}
\setlength{\footnotesep}{0.25cm}
\setlength{\jot}{10pt}
\titlespacing*{\section}{0pt}{4pt}{4pt}
\titlespacing*{\subsection}{0pt}{15pt}{1pt}
%%%END OF PAGE SETUP%%%

%%%FOOTER%%%
\fancyfoot{}
%\fancyfoot[LO,RE]{\vspace{-7.1pt}\includegraphics[height=9pt]{head_foot/LF}}
%\fancyfoot[CO]{\vspace{-7.1pt}\hspace{13.2cm}\includegraphics{head_foot/RF}}
%\fancyfoot[CE]{\vspace{-7.2pt}\hspace{-14.2cm}\includegraphics{head_foot/RF}}
\fancyfoot[RO]{\footnotesize{\sffamily{1--\pageref{LastPage} ~\textbar  \hspace{2pt}\thepage}}}
\fancyfoot[LE]{\footnotesize{\sffamily{\thepage~\textbar\hspace{3.45cm} 1--\pageref{LastPage}}}}
\fancyhead{}
\renewcommand{\headrulewidth}{0pt} 
\renewcommand{\footrulewidth}{0pt}
\setlength{\arrayrulewidth}{1pt}
\setlength{\columnsep}{6.5mm}
\setlength\bibsep{1pt}
%%%END OF FOOTER%%%

%%%FIGURE SETUP - please do not change any commands within this section%%%
\makeatletter 
\newlength{\figrulesep} 
\setlength{\figrulesep}{0.5\textfloatsep} 

\newcommand{\topfigrule}{\vspace*{-1pt}% 
\noindent{\color{cream}\rule[-\figrulesep]{\columnwidth}{1.5pt}} }

\newcommand{\botfigrule}{\vspace*{-2pt}% 
\noindent{\color{cream}\rule[\figrulesep]{\columnwidth}{1.5pt}} }

\newcommand{\dblfigrule}{\vspace*{-1pt}% 
\noindent{\color{cream}\rule[-\figrulesep]{\textwidth}{1.5pt}} }

\makeatother
%%%END OF FIGURE SETUP%%%

%%%TITLE, AUTHORS AND ABSTRACT%%%
\twocolumn[
  \begin{@twocolumnfalse}
{
\hfill\raisebox{0pt}[0pt][0pt]{
}\\[1ex]
}
\par
\vspace{1em}
\sffamily
\begin{tabular}{m{2.2cm} p{13.5cm} }

& \noindent\LARGE{\textbf{Empirical 
LiK excited state potentials: connecting short range and near dissociation 
expansions}}\\
\vspace{0.3cm} & \vspace{0.3cm} \\
 & \noindent\large{Sofia Botsi,\textit{$^{a}$} Anbang 
 Yang,\textit{$^{a}$} Mark M. Lam,\textit{$^{a}$} Sambit B. Pal,\textit{$^{a}$} 
 Sunil Kumar,\textit{$^{a}$} Markus Debatin,\textit{$^{a}$} and Kai 
 Dieckmann$^{\ast}$\textit{$^{a,b}$}} \\%Author names go here instead of "Full 
 %name", etc.
 
%TC:break Abstract
%\includegraphics{head_foot/dates}
& 
\noindent\normalsize{We report on a 
high\hyp{}resolution spectroscopic survey of 
${}^{6}\textrm{Li}{}^{40}\textrm{K}$ molecules near the 
$2\textrm{S}+4\textrm{P}$ dissociation threshold and produce a fully empirical 
representation for the $\textrm{B}^{1}\Pi$ potential by connecting available 
short\hyp{} and long\hyp{}range data. The purpose is to identify a suitable 
intermediate state for a coherent Raman transfer to the absolute ground state, 
and the creation of a molecular gas with dipolar interactions. Starting from 
weakly bound ultracold Feshbach molecules, the transition frequencies to  
twenty\hyp{}six vibrational states are determined. Our data are combined with 
long\hyp{}range measurements [Ridinger \textit{et al., EPL}, 2011, \textbf{96}, 
33001], and near\hyp{}dissociation expansions for the spin\hyp{}orbit coupled 
potentials are fitted to extract the $C_6$ dispersion coefficients. A suitable 
vibrational level is identified by resolving its Zeeman structure and by 
comparing the experimentally attained g\hyp{}factor to our theoretical 
prediction. Using mass\hyp{}scaling of the short\hyp{}range data for the 
$\textrm{B}^{1}\Pi$ [Pashov \textit{et al., Chem. Phys. Lett.}, 1998, 
\textbf{292}, 615\hyp{}620] and an updated value for its depth, we model the 
short\hyp{} and the long\hyp{}range data simultaneously and produce a 
Rydberg\hyp{}Klein\hyp{}Rees curve covering the entire range.} \\
%The abstract should be a single paragraph which summarises the content of the 
%article. Any references in the abstract should be written out in full 
%\textit{e.g.}\ [Surname \textit{et al., Journal Title}, 2000, \textbf{35}, 
%3523].
\end{tabular}

 \end{@twocolumnfalse} \vspace{0.6cm}
]
%%%END OF TITLE, AUTHORS AND ABSTRACT%%%

%%%FONT SETUP - please do not change any commands within this section
\renewcommand*\rmdefault{bch}\normalfont\upshape
\rmfamily
\section*{}
\vspace{-1cm}

%%%FOOTNOTES%%%
\footnotetext{\textit{$\ast$~Corresponding author E-mail: phydk@nus.edu.sg}}
\footnotetext{\textit{$^{a}$~Centre for Quantum Technologies (CQT), 3 Science Drive 2, Singapore 117543}}
%Address, Address, Town, Country. Fax: XX XXXX XXXX; Tel: XX XXXX XXXX; E-mail: xxxx@aaa.bbb.ccc
\footnotetext{\textit{$^{b}$~Department of Physics, National University of Singapore, 2 Science Drive 
3, Singapore 117542.}}%Address, Address, Town, Country.

%Please use \dag to cite the ESI in the main text of the article.
%If you article does not have ESI please remove the the \dag symbol from the title and the footnotetext below.
%\footnotetext{\dag~Electronic Supplementary Information (ESI) available: [details of any supplementary information available should be included here]. See DOI: 00.0000/00000000.}
%additional addresses can be cited as above using the lower-case letters, c, d, e... If all authors are from the same address, no letter is required

%\footnotetext{\ddag~Additional footnotes to the title and authors can be included \textit{e.g.}\ `Present address:' or `These authors contributed equally to this work' as above using the symbols: \ddag, \textsection, and \P. Please place the appropriate symbol next to the author's name and include a \texttt{\textbackslash footnotetext} entry in the the correct place in the list.}

%%%END OF FOOTNOTES%%%

%%%MAIN TEXT%%%%
%TC:break Main
\section{Introduction}
\label{Sec:intro}
Ultracold dipolar molecules have long been in the focus of experimental and 
theoretical research due to their long\hyp{}range and anisotropic interaction 
\cite{Quemener2012,Baranov2012,Bohn2017}. They provide a highly sensitive and 
robust platform for exploring the areas of quantum information processing 
\cite{deMille2002,Yelin2006,Ni2018,Hughes2020} and quantum simulation of 
long\hyp{}range spin models \cite{Micheli2006,Buchler2007,Yao2018}. They may as 
well play a significant role in precision measurements 
\cite{Andreev2018,Borkowski2018,Borkowski2019}, research on ultracold chemistry 
\cite{Carr2009,Balakrishnan2016,Yang2019} and as recently proposed in probing 
new physics beyond the Standard Model \cite{Safronova2018,Cairncross2019}. 
Their rich internal structure and molecular complexity however, renders their 
creation and coherent control a challenge. Thus far, a variety of 
bi\hyp{}alkali dimers have been produced in their absolute singlet 
ro\hyp{}vibronic ground state 
\cite{Ni2008,Park2015,Takekoshi2014,Molony2014,Guo2016,Voges2020,Cairncross2021}
by utlizing the coherent transfer scheme of stimulated Raman adiabatic passage 
(STIRAP) \cite{Vitanov2017}, in which the high initial molecular 
phase\hyp{}space density is preserved. The remarkable achievement of molecular 
quantum degeneracy has been achieved only for the case of 
${}^{40}\textrm{K}{}^{87}\textrm{Rb}$ \cite{DeMarco2019}. Alternative 
production approaches include direct laser cooling of the sample from a buffer 
gas source \cite{Anderegg2018}, and individual control of heavy neutral 
molecules in optical tweezers \cite{Anderegg2019,He2020}. Regarding the 
traditional three\hyp{}level STIRAP scheme, obtaining the desired efficient 
ground state transfer necessitates a detailed understanding of the molecular 
structure and an extensive spectroscopic survey for the identification of a 
suitable electronically excited state \cite{Park2015b,Zhu2016,Rvachov2018}. 
Selection rules for electronic transitions, Franck\hyp{}Condon overlap factors, 
mixing mechanisms between intermediate states and tuning capabilities of the 
available resources, are amongst some of the factors that need to be considered 
for making such a selection.

In this paper, we present results from the spectroscopic investigation of such 
intermediate candidate states, which is motivated by the objective to transfer ${}^{6}\textrm{Li}{}^{40}\textrm{K}$ molecules to the ground state. The long\hyp{}range part of the low\hyp{}lying $\textrm{B}^{1}\Pi$ was discussed in our previous work \cite{Brachmann2012} as a possible candidate, and a specific vibrational sub\hyp{}level was selected. 
Here, we show the long\hyp{}range spectrum below the ${}^{6}$Li($2^2$S$_{1/2}$)+${}^{40}$K($4^2$P$_{3/2}$) 
asymptote and the line assignment analysis. In order to understand the spectral 
structure, we explore the intermediate state mixing due to the spin\hyp{}orbit 
coupling interaction. The Zeeman sub\hyp{}structure of the selected vibrational 
level is resolved and the experimentally attained g\hyp{}factor is compared to 
our theoretical prediction. To access the levels of the $\textrm{B}^{1}\Pi$ 
potential, we associate ultracold ${}^{6}\textrm{Li}$ and ${}^{40}\textrm{K}$ 
atoms via a magnetically tunable Feshbach resonance \cite{Koehler2006} and 
apply spectroscopic light. This scheme differs from previous studies, which 
were performed by conventional photoassociation (PA) in a dual\hyp{}species 
magneto\hyp{}optical trap (MOT) \cite{RidingerPA2011} and by Doppler\hyp{}free 
polarization labelling spectroscopy (PLS) of the 
${}^{7}\textrm{Li}{}^{39}\textrm{K}$ isotopologue in a heat\hyp{}pipe 
\cite{Pashov1998}. They provide important information on a wide range of the 
excited spectrum, but they do not cover all the levels in the region of our 
interest. The PA results when combined with our data, apart from facilitating 
the level assignment, assist in inferring the $C_6$ parameters. By combining 
them with the PLS observations, we produce a complete empirical 
Rydberg\hyp{}Klein\hyp{}Rees (RKR) \cite{Rydberg1933,Klein1932,Rees1947} curve 
for the $\textrm{B}^{1}\Pi$ potential.   

\section{Spectroscopic results and line assignment}
\label{Sec:spectroscopy}
%%% Our system -----------------------------------------------------------------
The starting point of our experiments is the creation of a quantum degenerate 
mixture of $10^{5}$ ${}^{6}\textrm{Li}$ and $8\times 10^{4}$ 
${}^{40}\textrm{K}$ atoms in a magnetic trap, which is sympathetically cooled 
via evaporative cooling of bosonic ${}^{87}\textrm{Rb}$ \cite{Taglieber2008}. 
The Fermi\hyp{}Fermi mixture is then transferred into a crossed optical dipole 
trap, where ${}^{6}\textrm{Li}$ and ${}^{40}\textrm{K}$ atoms are prepared in 
the $\ket{F_{\mathrm{Li}}=1/2,m_{F,\mathrm{Li}}=-1/2}$ and 
$\ket{F_{\mathrm{K}}=9/2,m_{F,\mathrm{K}}=-9/2}$ hyperfine states. Here, $F$ is 
the hyperfine quantum number and $m_{F}$ its respective projection along the 
internuclear axis. Magneto\hyp{}association is performed by sweeping the 
magnetic field across an interspecies Feshbach resonance located at 
$21.56\,$mT, which results in up to $10^{4}$ Feshbach molecules 
\cite{Voigt2009}. The molecular state contains a significant admixture from the 
singlet ground state potential \cite{Yang2020} and is an excellent starting 
point for the excited state spectroscopy. Uncombined free atoms are spatially 
separated from weakly bound molecules by means of an inhomogeneous 
magnetic\hyp{}field pulse, as the latter possess an almost vanishing magnetic 
moment. The pulse is applied during time\hyp{}of\hyp{}flight (TOF) after 
release from the trap, and is followed by detection via absorption imaging.

%%% One-photon spectroscopy ----------------------------------------------------
The ultracold mixture is illuminated by spectroscopic light and 
one\hyp{}photon spectroscopy is performed for the investigation of the 
vibrational levels of the excited potentials during TOF before imaging. If the 
spectroscopic light is resonant with an electronically excited state, then the 
molecules undergo resonant excitation and subsequent spontaneous decay, which 
is highly likely to occur to some other molecular bound state of the ground 
state and not to the initial Feshbach state. This process will manifest as a 
loss in the number of detected molecules during absorption imaging of the 
Feshbach state. Since this scheme is destructive, a new molecular sample is 
prepared after each experimental cycle. The spectroscopic source is a 
commercial external cavity diode laser (Toptica DL Pro), which is tunable over 
a broad wavelength range of $760\,$nm to $775\,$nm and has a nominal output 
power of $28\,$mW. The laser's frequency is measured by an optical beat note 
with a frequency comb (FC) that is deriving its long\hyp{}term stability from a 
GPS\hyp{}disciplined RF reference generator. To determine the frequency of the 
laser to within one free spectral range of the FC, a home\hyp{}built wavemeter 
is utilized. It has an accuracy of $10\,$MHz, which is accomplished by 
referencing it to a laser locked to the potassium D2\hyp{}line.

{ % begin box to localize effect of arraystretch change
	\renewcommand{\arraystretch}{1.5}
	\setlength{\arrayrulewidth}{0.1pt} 
\begin{table}[ht]
\small
%Tables typeset in RSC house style do not include vertical lines. Table 
%footnote symbols are lower-case italic letters and are typeset at the bottom 
%of the table. Table captions do not end in a full point
\caption{\ Measured long\hyp{}range vibrational levels close to the 
		2S+4P asymptote. Counting is downwards from the dissociation threshold. 
		The transition frequencies are computed with respect to the hyperfine 
		free ground state asymptote. The frequency detunings utilized for 
		obtaining the $C_6$ coefficients are referenced to the same hyperfine 
		state of the excited state asymptote as the one used in 
		\cite{RidingerPhD2011}}	
\label{tbl:measuredlines}
	\begin{tabular*}{0.48\textwidth}{@{\extracolsep{\fill}}lllll}
		\hline
		state & group & $v$ & f$_{\textrm{laser}}$ (THz) & $\Delta$f (GHz) \\
		\hline
		$\Omega$=1$^{\textrm{up}}$  &dyad 	     &-6  & 390.474969 & \, 544.52\\
									&  		     &-7  & 390.154969 & \, 864.52\\
									& 	    	 &-8  & 389.734896 & 1284.32\\
								   	&	    	 &-9  & 389.215496 & 1803.72\\
									&	    	 &-10 & 388.611196 & 2408.02\\
									&	    	 &-11 & 387.925196 & 3094.02\\
									&	    	 &-12 & 387.125196 & 3894.02\\
		\hline
		$\Omega$=0$^{\textrm{+}}$   &upper triad &-7  & 390.606196 & \, 413.02\\
									&			 &-8  & 390.396196 & \, 623.02\\
									&			 &-9  & 390.129196 & \, 890.02\\
									&			 &-10 & 389.781196 & 1238.02\\
									&			 &-11 & 389.347196 & 1672.02\\ 
									&			 &-12 &	388.865196 & 2154.02\\  
		\hline     
		$\Omega$=0$^{\textrm{-}}$   &upper triad &-7  & 390.577196 & \, 442.02\\
									&			 &-8  & 390.356196 & \, 663.02\\
									&			 &-9  & 390.082196 & \, 937.02\\
									&			 &-10 & 389.737196 & 1282.02\\
	   								&			 &-11 & 389.337196 & 1682.02\\ 
		\hline
		$\Omega$=1$^{\textrm{down}}$&upper triad &-7  & 390.521196 & \, 498.02\\
		\hline
		$\Omega$=0$^{\textrm{+}}$   &lower triad &-4  & 389.171196 & \, 118.52\\
									&			 &-5  & 389.037196 & \, 251.90\\
									&			 &-6  & 388.855196 & \, 433.90\\
									&			 &-7  & 388.621196 & \, 667.90\\
		\hline
		$\Omega$=1                  &lower triad &-3  & 389.215496 & \, \, 
		73.60\\
									&			 &-4  & 389.109961 & \, 179.90\\
									&			 &-5  & 388.925196 & \, 363.90\\
	\end{tabular*}
\end{table}
} % end box

%%% Line assignments by comparison to PA data ----------------------------------
In Table~\ref{tbl:measuredlines} we present a summary of the measured 
long\hyp{}range states located up to $4\,$THz below the 
${}^{6}$Li($2^2$S$_{1/2}$)+${}^{40}$K($4^2$P$_{3/2}$) asymptote. To 
enable a broad survey, the spectroscopic resolution of the measurement is 
initially set to $1\,$GHz. This is sufficient to unambiguously identify the 
states, since the level spacing between adjacent vibrational levels is much 
larger. The experimentally observed transitions are assigned to the nearest 
predicted level based on extrapolation of the PA lines, for which the 
assignment was done based on progressions described by the LeRoy\hyp{}Bernstein 
law \cite{RidingerPA2011} as further described in Section~\ref{Sec:NDEC6}. Six 
vibrational series are distinguished from each other which contain a total of 
$26$ vibrational levels. The long\hyp{}range potentials are coupled by the 
strong spin\hyp{}orbit interaction and are labelled with the quantum number 
$\Omega^{\pm}$, as is suitable for Hund's case (c) molecules. $\Omega$ is the 
projection of the coupled angular momentum $\textbf{J}_a=\textbf{S}+\textbf{L}$ 
on the molecular axis, where $S$ is the total spin and $L$ is the orbital 
angular momentum. The $\pm$ denotes the reflection symmetry of the spatial 
component of the  electronic wave function through a plane containing the 
internuclear axis and the superscripts up/down further classify the 
long\hyp{}range states into groups of potentials for unambiguous distinction. 
The frequency detunings are computed by subtracting from our measured 
transition frequencies the energy of the Feshbach molecular state with respect 
to the hyperfine\hyp{}free ground state asymptote 
${}^{6}$Li($2^2$S$_{1/2}$)+${}^{40}$K($4^2$P$_{3/2}$). This energy contains the 
binding 
\begin{figure}[ht]
	\centering
	\includegraphics[width=8.5cm]{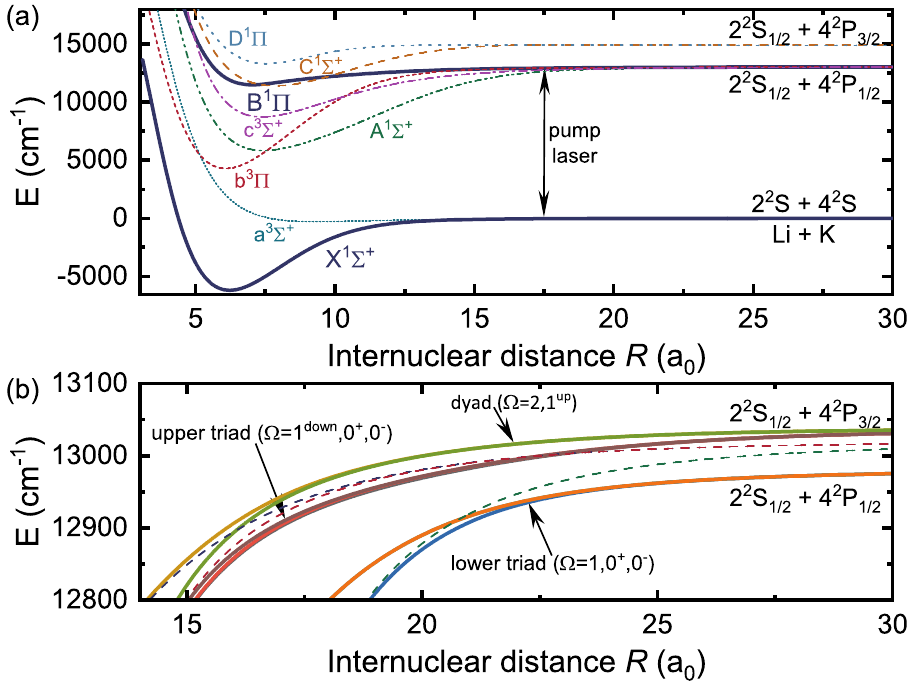}
	\caption{Potential energy curves of ${}^{6}\textrm{Li}{}^{40}\textrm{K}$  	
		molecules. (a) Adiabatic potentials, with no spin-orbit coupling 
		effects considered, connecting to the lowest three electronic 
		asymptotes. The ground state potential is given by Tiemann et al. 
		\cite{Tiemann2009} and the excited state potentials by Allouche 
		\cite{Allouche}. (b) Spin\hyp{}orbit coupled potentials near the 
		dissociation threshold. The dyad ($\Omega=2,1^{\textrm{up}}$), upper 
		triad ($\Omega=0^+,0^-,1^{\textrm{down}}$) and lower triad 
		($\Omega=1,0^+,0^-$) are labelled. Bare potentials are included with 
		dashed lines to show the rapidly diminishing perturbation caused by the 
		spin\hyp{}orbit coupling as deeper bound levels are considered}
	\label{fgr:AllPotCombined}
\end{figure} 
energy of the molecular state and its Zeeman shift at the Feshbach magnetic 
field of $21.56\,$mT ($472.5\,$MHz), and the atomic hyperfine energies of 
${}^{6}\textrm{Li}$ ($152.1\,$MHz) and ${}^{40}\textrm{K}$ ($571.5\,$MHz) for 
the respective asymptotic states.

%%% SOC potentials -------------------------------------------------------------
Further characterization of the long\hyp{}range states requires a detailed 
study of the spin\hyp{}orbit coupling interaction, which leads to the mixing of 
neighboring singlet and triplet potentials and becomes dominant at large 
internuclear distances. Here, the relevant excited short\hyp{}range curves 
which result into the coupled long\hyp{}range potentials are the 
$\textrm{B}^{1}\Pi$, $\textrm{b}^{3}\Pi$, $\textrm{A}^{1}\Sigma^{+}$ and the 
$\textrm{c}^{3}\Sigma^{+}$ as shown in Fig.~\ref{fgr:AllPotCombined}(a), where 
the $\textrm{A}^{1}\Sigma^{+}$ and $\textrm{b}^{3}\Pi$ cross at an internuclear 
distance of $7.5\,\textrm{a}_0$, as is commonly observed in alkali dimers. 
Fig.~\ref{fgr:AllPotCombined}(b) shows the region of strongest spin\hyp{}orbit 
coupling, where the eight Hund’s case (c) long\hyp{}range states dissociate to 
both of the $2\textrm{S}+4\textrm{P}$ asymptotes of the 
${}^{6}\textrm{Li}{}^{40}\textrm{K}$ molecule. The singlet\hyp{}triplet mixing 
is calculated by projecting the spin\hyp{}orbit coupled states onto the bare 
potential basis. This is necessary to facilitate the selection of a suitable 
intermediate state that will mediate coupling between the dominantly singlet 
Feshbach molecular state and the singlet ground state for the two\hyp{}photon 
transfer. The $\Omega=1^{\textrm{up}}$ state of the dyad and the $\Omega=0^-$ 
state of the upper triad meet this requirement, as they contain a large singlet 
component and connect to singlet bare potentials in the short\hyp{}range 
(Fig.~S1). Moreover, the $\Delta \Omega=0,\pm1$ and $\Delta 
S=0$ selection rules further narrow down the choice, making the 
$\Omega=1^{\textrm{up}}$ potential a promising option.
\section{Zeeman effect for Hund's case (c) molecules}
\label{Sec:HundC}
\begin{figure}[bh]
	\centering
	\includegraphics[width=8.5cm]{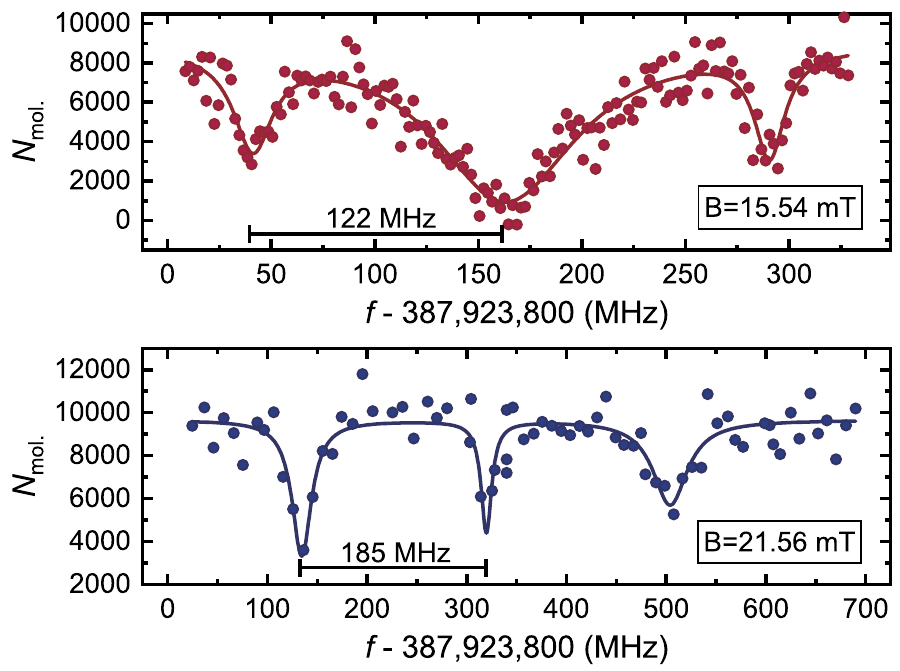}
	\caption{Zeeman triplets for the $v=-11$ level measured at two different 
		magnetic fields. The relative size of Zeeman components changes with 
		polarization. The Zeeman splittings correspond to a measured average 
		g-factor value, which is in good agreement with our theoretical 
		prediction}
	\label{fgr:Zeemantripletts}
\end{figure}

%%% Experimental results & g-factor calculation for Hund's case (c) ------------
To identify the $\Omega$=1$^\textrm{up}$ state as a suitable intermediate 
state, it is desirable to resolve its characteristic magnetic Zeeman structure. 
Here, only the rotational ground state is of interest, and 
therefore the total angular momentum is $J=1$. Three magnetic sublevels are 
expected, which are denoted by $M_J$. For this measurement scans with higher 
resolution are performed by employing an interferometric frequency 
stabilization device, which provides in\hyp{}lock frequency tuning of the laser 
in MHz steps over a large range \cite{Brachmann12}. In order to resolve the 
magnetic sub\hyp{}structure, we iteratively adjust the spectroscopy laser power 
to avoid power\hyp{}broadening and reduce the irradiation time. In 
Fig.~\ref{fgr:Zeemantripletts} the Zeeman triplet substructure of the $v=-11$ 
vibrational level is shown, where $v$ is the vibrational quantum number. It is 
measured at two different magnetic fields, specifically at the $15.54\,$mT and 
at the $21.56\,$mT Feshbach resonance of ${}^{6}\textrm{Li}{}^{40}\textrm{K}$. 
Zeeman splittings of $122\,$MHz and $185\,$MHz are observed, respectively. This 
is consistent with a linear Zeeman effect and an average g-factor of 
$g_{\textrm{exp.}}=0.59$. In the following, this value is compared to 
the theoretical prediction for the g-factor for the  Hund's case (c) vector 
coupling scheme, where we generally represent the vibrational levels in the 
coupled $\ket{v(\Omega),J_{a};\Omega,\epsilon, J, M_J}$ basis. Here, the total 
parity $\epsilon$ is specified, which is related to whether symmetric or 
anti\hyp{}symmetric combinations of $\pm\Omega$\hyp{}states are utilized. The 
effective Hamiltonian describing the interaction between the external magnetic 
field and electron spin and the orbital magnetic moments in spherical tensor 
notation is \cite{Brown2003}:
\begin{equation}
	\mathcal{H}_Z=g_s\mu_{\textrm{B}}\textrm{T}^1(\textbf{B})\cdot\textrm{T}^1(\textbf{S})+g_L\mu_{\textrm{B}}\textrm{T}^1(\textbf{B})\cdot\textrm{T}^1(\textbf{L})\quad ,
	\label{eqn:ZHamiltonian}
\end{equation}	
\noindent where $\mu_{\textrm{B}}$ is the Bohr magneton and $g_S$ and $g_L$ are 
the g\hyp{}factors for the electron spin and the orbital motion respectively. 
The effective g\hyp{}factor is then directly related to the expectation value 
via $\left<\mathcal{H}_Z\right>=M_J g_{\textrm{(c)}} \mu_{\textrm{B}} B$. In 
order to evaluate $\left<\mathcal{H}_Z\right>$, we follow the general scheme as 
exemplified in \cite{Carrington1995} that expands to the reduced matrix 
elements that can be evaluated with the help of the quantum numbers of the 
uncoupled basis. As the observed Zeeman shift is smaller than the rotational 
constants of the $v=-11$, off\hyp{}diagonal matrix element for different $J$ do 
not need to be considered. Then, for our case of rotationless excitation to the 
$\Omega=1^{\textrm{up}}$ state, we have $J_a=1$ and the Zeeman shift is the 
same for both parity eigenstates. As we work with the $2\textrm{S}+4\textrm{P}$ asymptote, $L=1$ is the sole contribution to the 
spin\hyp{}orbit coupled state. However, for the spin, the superposition of the 
$S=0$ and $S=1$ states contributing to the $\Omega=1^{\textrm{up}}$ state needs to be considered. From 
our analysis of the spin\hyp{}orbit coupling (Fig.~S1), we see that for the 
long\hyp{}range there is an equal admixture of the $\ket{^1\Pi}$ and 
$\ket{^3\Pi}$ components, whereas for the short\hyp{}range the state becomes 
purely of $\ket{^1\Pi}$ character. Hence, we adopt a simple approach by taking 
the average of the results for the g-factor between the short\hyp{}range and long\hyp{}range spin compositions. A more accurate calculation would require to 
integrate the spin composition weighted by the probability density of the corresponding vibrational wave function. Here, we find as a result a g-factor for the $\Omega$=1$^\textrm{up}$ state of 
$g_{\textrm{(c)}}=0.625$, which is in good agreement with the measured value. 
In comparison, for the Hund's case (a) the respective value is  $g_{\textrm{(a)}}=(\Lambda+2\Sigma)\Omega/J(J+1)=0.5$ \cite{Schadee1978}, 
where $\Lambda$ is the projection of $\textbf{L}$ along the intermolecular axis. Therefore, the measurements of the Zeeman effect support the validity of the 
Hund's case (c) coupling scheme as an appropriate description for vibrational 
levels as deeply bound as the $v=-11$.
\section{Near\hyp{}dissociation expansions and $C_6$ coefficients}
\label{Sec:NDEC6}
To achieve a more complete characterization of the long\hyp{}range behavior of 
the potentials, the data set based on PA measurements \cite{RidingerPA2011} is 
extended by our measurements of more deeply bound vibrational levels, as 
already mentioned in Section~\ref{Sec:spectroscopy}. This allows to determine 
the $C_6$ dispersion coefficients from a larger data set for each vibrational 
progression. Additionally, the description by the semi\hyp{}classical 
LeRoy\hyp{}Bernstein formula is extended and near\hyp{}dissociation expansions 
(NDE) are used \cite{LeRoy1970,LeRoy1970b}.

%%% Combination of our data with PA data ---------------------------------------
The PA measurements resulted in seven vibrational series below the 
${}^{6}\textrm{Li}(2^2$S$_{1/2}) + {}^{40}\textrm{K}(4^2$P$_{3/2})$ asymptote. 
In order to combine our data to the PA measurements, we compute the frequency 
detunings $\Delta$f (shown in Table~\ref{tbl:measuredlines}) with respect to 
the $^{40}\textrm{K}$ $4S_{1/2}(F=9/2)\rightarrow4P_{3/2}(F'=11/2)$ hyperfine 
transition frequency \cite{Falke2006}, which is used as a reference for the PA 
measurements \cite{RidingerPhD2011}. For the data comparison we assume the 
hyperfine\hyp{}free asymptotic energy of the $\textrm{X}^{1}\Sigma^{+}$ ground 
state as a reference point for our measurements. This reflects that the PA 
measurements were performed in a MOT, where for the initial states all four 
hyperfine ground states of $^{6}\textrm{Li}$ and $^{40}\textrm{K}$ are 
possible. Hence, the resulting frequency uncertainty in the data comparison is 
on the order of the hyperfine energies, which is comparable to the measurement 
resolution of $1\,$GHz. A more precise comparison would require hyperfine 
resolved measurements, which are difficult to be achieved for all the PA lines.

The general NDE expressions for the vibrational energies $G_v$ and the 
rotational constants $B_v$ are \cite{Tromp1985,Appadoo1996}:
\begin{equation}
\begin{aligned}
		G_v=\mathcal{D}-K_0^{\infty}(v)\cross\mathcal{F}_0(v_{\mathcal{D}}-v)\\ 
		B_v=K_1^{\infty}(v)\cross\mathcal{F}_1(v_{\mathcal{D}}-v)\quad ,
	\label{eqn:NDE}
\end{aligned}
\end{equation}
\noindent where $\nu_{\mathcal{D}}$ is the extrapolated non\hyp{}integer 
effective vibrational index at the dissociation energy $\mathcal{D}$. The 
$K_m^{\infty}(v)$ functions are:
 \begin{figure}[ht]
	\centering
	\includegraphics[width=8.5cm]{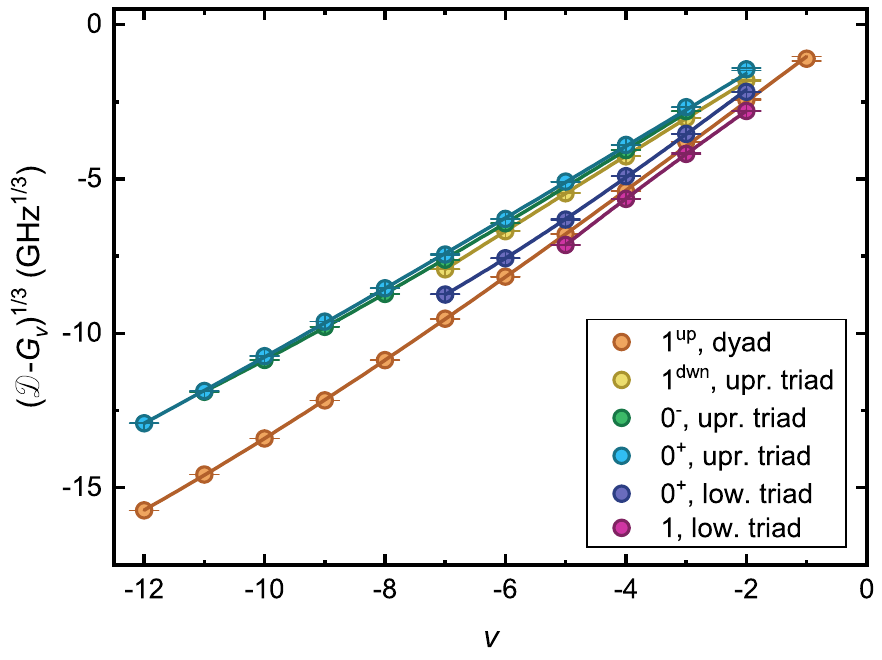}
	\caption{Cubic root of the binding energies plotted for various vibrational 
		levels of the long\hyp{}range potentials by combining our data with the 
		PA measurements. The states belonging to the upper\hyp{} and the lower 
		triad are grouped together since they dissociate to the same asymptote}
	\label{fgr:Gv}
\end{figure}
\begin{equation}
	K_m^{\infty}(v)=X_m(n,C_n,\mu)(v_{\mathcal{D}}-v)^{[2n/(n-2)]-2m}\quad ,
\end{equation}
\noindent where $X_m(n,C_n,\mu)=\bar{X}_m(n)/[\mu^nC_n^2]^{1/n-2}$ are 
numerical factors depending on physical constants \cite{LeRoy1972} and $\mu$ is 
the Watson's charge\hyp{}modified reduced mass. The empirically determined 
functions $\mathcal{F}_m(v_{\mathcal{D}}-v)$ that are required to approach 
unity close to the dissociation threshold are expressed in the form of a Pad\'e 
expansion using rational polynomials \cite{Goscinski1972}:
{ % begin box to localize effect of arraystretch change
	\renewcommand{\arraystretch}{1.5}
	\begin{table}[b]
		\small
		\centering
		\caption{\ Values of the $C_6$, $v_{\mathcal{D}}$ and $p_2^0$ fitting 
			parameters and their respective errors $\delta C_6$,
			$\delta$v$_\mathcal{D}$ and $\delta p_2^0$ as obtained from fitting 
			to a NDE with $S=1$. The expansion to which the parameters apply is given in parentheses. The values of the $p_2^0$ parameters are 
			multiplied by $10^2$. The $C_6$ parameters are given in atomic 
			units}
		\label{tbl:GvFittingParam}
		\begin{tabular*}{0.48\textwidth}{@{\extracolsep{\fill}}ccccc}
			\hline
			%\hline
			$\Omega$ & $C_6$ $\pm$ $\delta C_6$ & $v_\mathcal{D}$ $\pm$ 
			$\delta v_\mathcal{D}$ & $p_2^0\pm\delta p_2^0$ & 
			$p_2^0\pm\delta p_2^0$ \\
			& ($G_v, B_v$) & ($G_v, B_v$) & ($G_v$) & ($B_v$)\\
			\hline
			1$^{\textrm{up}}$   & 8619 $\pm$ 736 & -0.29 $\pm$ 0.07 
			& -0.16 $\pm$ 0.02 & 0.16 $\pm$ 0.72\\
			0$^{\textrm{+}}$    & 30391 $\pm$ 4984 & -0.65 $\pm$ 0.13 & 
			-0.08 $\pm$ 0.04 & 1.35 $\pm$ 1.01\\ 
			0$^{\textrm{-}}$    & 24880 $\pm$ 3016 & -0.65 $\pm$ 0.10 & 
			-0.16 $\pm$ 0.03 & 0.43 $\pm$ 0.88\\
			1$^{\textrm{down}}$ & 27717 $\pm$ 8474 & -0.46 $\pm$ 0.16 & 
			$\,$ 0.08 $\pm$ 0.22 & 0.81 $\pm$ 0.73\\
			0$^{\textrm{+}}$    & \ \ 8251 $\pm$ 2169 & -0.55 $\pm$ 
			0.13 & -0.50 $\pm$ 0.18 & 6.68 $\pm$ 5.02\\
			1                   & 13309 $\pm$ 9098 & $\,$ 0.05 $\pm$ 
			0.24 & \ 0.54 $\pm$ 0.89 & 
		\end{tabular*}
	\end{table}
} % end box
\begin{equation}
	\mathcal{F}_m(v_{\mathcal{D}}-v)=\Bigg(\frac{1+\sum_{i=t}^{L}p_i^m(v_{\mathcal{D}}-v)^i}{1+\sum_{j=t}^{M}q_j^m(v_{\mathcal{D}}-v)^j}\Bigg)^S\quad
	 ,
	\label{eqn:Expansion}
\end{equation}
\noindent where the power of the exponent $S$ is set at either $S=1$ to yield 
an \mbox{''}outer\mbox{''} expansion, or at $S=2n/(n-2)$, to yield an 
\mbox{''}inner\mbox{''} expansion. The $p_i^m$ and $q_j^m$ are the parameters 
of the expansion. For the case of the leading terms of the attractive 
long\hyp{}range potentials having powers of $n=6$ or $n=8$, $t=1$ is applicable 
\cite{LeRoy1980}.
 
The measured lines are shown in Fig.~\ref{fgr:Gv}, where the cubic root of the 
vibrational energies relative to the excited state asymptote is plotted versus 
the vibrational index $v$ for the extended data set. For each long\hyp{}range 
potential, the $C_6$ coefficients, the $v_{\mathcal{D}}$ values, as well as the 
expansion parameters are extracted from the fitting and are listed in 
Table~\ref{tbl:GvFittingParam} along with their error estimates. An 
\mbox{''}outer\mbox{''} expansion is performed for all of the $\Omega$ states. 
Various combinations of the $p_i^m$ and $q_j^m$ expansion parameters extended 
to different orders are tested for each long\hyp{}range potential and the 
fitting quality is assessed. To avoid large estimation errors due to a large 
number of fitting parameters, a second order expansion using only $p_2^0$ is 
performed for all the excited states. It should be noted here that this Pad\'e 
analysis directly corresponds to using the improved LeRoy-Bernstein NDE formula 
\cite{Comparat2004} for the case of $n=6$. The latter makes use of a quadratic 
term as the leading order beyond the pure $C_6$ semiclassical 
LeRoy\hyp{}Bernstein formula. 

%%% C_6 results discussion -----------------------------------------------------
Most of our extracted $C_6$ values agree within 10\% with the PA results and 
the respective theoretical  predictions \cite{Bussery1987}. A slightly higher 
deviation is observed for the $\Omega$=1$^{\textrm{up}}$. For this state the 
modified LeRoy\hyp{}Bernstein radius \cite{Bing1995} is 
$R_{\textrm{mLR}}=19.8\,$a$_0$, while the deepest bound vibrational level 
reached by our measurements possesses a classical outer turning point at 
$15.6\,$a$_0$, as inferred from the RKR analysis presented in the next section. 
Similarly for the $\Omega$=0$^{\textrm{+}}$ state of the lower triad 
$R_{\textrm{mLR}}=29.5\,$a$_0$, while the deepest measured vibrational level 
has its turning point at $20.8\,$a$_0$. Nevertheless, we observe that the $C_6$ 
values remain stable, when varying the extend of the data set to include only 
less deeply bound vibrational levels. In contrast, when a pure $C_6$ expansion 
is utilized, the resulting dispersion coefficient varies strongly with the 
extend of the used data set. Hence, a NDE method is clearly more appropriate than a 
pure $C_6$ semiclassical LeRoy\hyp{}Bernstein formula. For the two 
long\hyp{}range states belonging to the lower triad, the contribution of the 
next higher order $C_8/R^8$ term in the multipolar expansion of the interaction 
potential needs to be considered. At intermolecular distances of $64.4\,$a$_0$ 
and $50.7\,$a$_0$ for the $\Omega$=0$^{\textrm{+}}$ and the $\Omega$=1 state 
respectively, which are well within our spectroscopic reach, the contributions 
of the $C_8$ coefficients become significant and need to be included in the NDE 
formula. However, it has been suggested \cite{Zhu2016,Li2019} that the precise 
value of the C$_8$ is hard to obtain accurately when fitting with the improved 
LeRoy\hyp{}Bernstein NDE expression. Due to the limited number of measured 
lines for these states, accurate modeling with a NDE including a larger number 
of fitting parameters is not feasible. 

In the short\hyp{}range a large data basis is available for rotationally 
excited states from the PLS measurements. For the purpose of extending the 
description of rotationally excited states to the long\hyp{}range, a NDE for 
the rotational energies $B_v$ is fitted with the respective expression 
introduced in eqn~(\ref{eqn:NDE}). However, for this purpose only a small data 
set from the PA measurements is available, since our spectroscopy does not 
cover rotationally excited states. A Pad\'e expansion with one $p_2^0$ 
parameter is utilized, where the $C_6$ and $v_{\mathcal{D}}$ parameters are 
taken from our $G_v$ fits. The results are included in 
Table~\ref{tbl:GvFittingParam}. 

\section{Combining short\hyp{} and long\hyp{}range data}
\label{Sec:Combining}
Thus far, merging our measurements with the PA observations yields an extended 
characterization of the spin\hyp{}orbit coupled states near the threshold. This 
holds in particular for the $\Omega$=1$^{\textrm{up}}$, which is of interest as 
an intermediate state for the two\hyp{}photon transfer of the Feshbach  
molecules to the dipolar ground state. This state connects to the 
$\textrm{B}^{1}\Pi$ potential in the short\hyp{}range (Fig.~S1). At the 
inner turning point high\hyp{}lying vibrational levels of this potential have a 
large Franck\hyp{}Condon overlap with the absolute ground state, favoring large 
transitions strengths. For the $\textrm{B}^{1}\Pi$ potential a large data set 
exists in the short\hyp{}range for the ${}^{7}\textrm{Li}{}^{39}\textrm{K}$ 
isotopologue measured by PLS \cite{Pashov1998}. Here, we combine the 
short\hyp{}range and long\hyp{}range data to attain an improved potential 
curve. To our knowledge it is unique that an excited potential is supported by 
empirical data throughout almost the entire internuclear range. The 
short\hyp{}range data cover the range from the $v=-33$ vibrational ground state 
up to the $v=-15$ level. Our measurements cover the $v=-12$ to $v=-6$ states, 
whereas the PA data range from $v=-5$ to the $v=-1$. From our analysis it is 
apparent that there are no available experimental results for the $v=-14$ and 
$v=-13$ levels. 

%%% Mass scaling of data from PLS in the short range ---------------------------
To facilitate the combination of the data we use mass\hyp{}scaling of the Dunham coefficients determined by the PLS measurements to our ${}^{6}\textrm{Li}{}^{40}\textrm{K}$ isotopologue by the ratio of the reduced masses $\widetilde{\mu}=\mu\left({}^{7}\textrm{Li}{}^{39}\textrm{K}\right)/\mu\left({}^{6}\textrm{Li}{}^{40}\textrm{K}\right)$. The vibrational term energies obtained from the Dunham expansion are:
\begin{equation}
	T(\tilde v,J)=T_\mathrm{e}({\textrm{B}^1\Pi})+\sum_{k,l}\widetilde{\mu}^{(l+k)/2}Y_{k,l}(\tilde v+\frac{1}{2})^k(J(J+1))^l\quad ,
\end{equation}
\noindent where $Y_{k,l}$ are the Dunham coefficients, $\tilde v$ is the 
vibrational level indexed by positive numbers starting from $\tilde v=0$ for 
the ground state. As the PLS data are obtained by measuring transitions 
originating from low\hyp{}lying states of the $\textrm{X}^{1}\Sigma^{+}$, the 
term energy for the excited state potential $T_\mathrm{e}({\textrm{B}^1\Pi})$ 
is defined relative to the minimum of the ground state potential 
\cite{Pashov1998}. In the same work the short\hyp{}range data are extrapolated 
to obtain the asymptotic energy of the $\textrm{B}^1\Pi$ potential and hence 
infer the potential depth $\mathcal{D}({\textrm{B}^1\Pi})$. Here, we calculate 
the depth in a different way to combine the short\hyp{}range data with the 
long\hyp{}range data, both referenced to the $\textrm{B}^1\Pi$ asymptote. From 
our previous high\hyp{}resolution two\hyp{}photon spectroscopy of the 
$\textrm{X}^{1}\Sigma^{+}$ ground state \cite{Yang2020} the depth of the ground 
state potential was accurately measured as 
$\mathcal{D}(\textrm{X}^{1}\Sigma^{+})=6216.863166\,$cm$^{-1}$. Further, the 
wavenumber of the D2\hyp{}potassium atomic transition
$\bar{\lambda}_{\mathrm{D2}}=13042.899700\,$cm$^{-1}$ is known from literature 
\cite{Falke2006}. Therefore, an improved value for the depth of the excited 
state potential can be inferred from the measured data as:
\begin{equation}
		\mathcal{D}({\textrm{B}^1\Pi})=\mathcal{D}(\textrm{X}^{1}\Sigma^{+})+\bar{\lambda}_{\mathrm{D2}}-T_\mathrm{e}({\textrm{B}^1\Pi})=1687.002866\,\textrm{cm}^{-1}\quad
		 .
\end{equation}
We produce an updated semiclassical RKR potential curve for the ${\textrm{B}^1\Pi}$ using the RKR1 program by LeRoy \cite{LeRoy2017}. The RKR1 
program allows for a mixed representation of the ro-vibrational energies by the 
mass\hyp{}scaled Dunham parameters $Y_{k,l}$ for the short\hyp{}range, and 
simultaneously by the near\hyp{}dissociation expansion parameters $C_6$, 
$v_{\mathcal{D}}$ and $p_2^0$ (vibrational and rotational) 
\begin{figure}[hb]
	\centering
	\includegraphics[width=8.5cm]{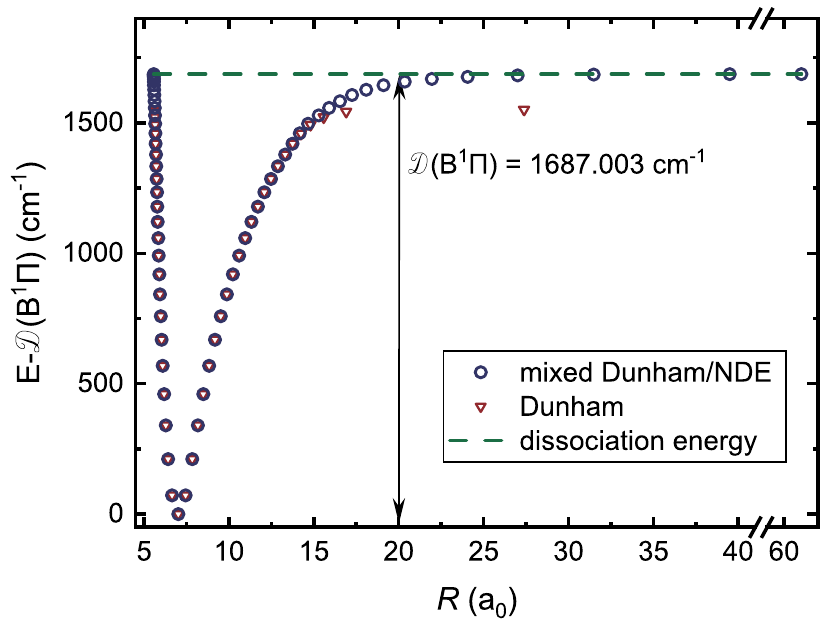}
	\caption{Complete empirical RKR curve for the excited $\textrm{B}^{1}\Pi$ 
	potential. A mixed Dunham/NDE representation is utilized for the 
	vibrational energies (blue circles). A pure Dunham representation (red 
	triangles) fails to accurately account for the shape of the potential near 
	the outer turning points}
	\label{fgr:RKRpotential}
\end{figure}
for the long\hyp{}range. We make use of $\mathcal{D}({\textrm{B}^1\Pi})$ to relate the 
two energy scales. The RKR1 program interpolates between the two ranges by utilizing a switching function 
$F_{\textrm{s}}(v)=\bigl\{1+exp\big((v-v_\textrm{s})/\delta 
v_\textrm{s}\big)\bigr\}^{-1}$. We use $v_\textrm{s}=-14$ and $\delta 
v_\textrm{s}=0.5$ for the switch\hyp{}over point and range, which is 
conveniently located at the small gap of available vibrational experimental 
data. The resulting RKR potential curve is shown in Fig.~\ref{fgr:RKRpotential} 
represented by the inner and outer classical turning points for each 
vibrational level. The corresponding numerical values are tabulated in 
Table~S2. For comparison a potential curve only based on the Dunham 
representation is plotted as well. As seen in Fig.~\ref{fgr:RKRpotential}, the 
Dunham curve fails to correctly represent the limiting near\hyp{}dissociation 
behavior of the vibrational energies as expected. 
%Smoothing of the inner turning points is performed for high vibrational energies where the repulsive inner wall of the potential is very steep and exhibits unphysical behaviour. The classical turning points of the RKR potential are presented in the Appendix.
%
\section{Conclusions}
To conclude, an extensive investigation of the vibrational states 
of ${}^{6}\textrm{Li}{}^{40}\textrm{K}$ molecules below the 
${}^{6}$Li($2^2$S$_{1/2}$)+${}^{40}$K($4^2$P$_{3/2}$) asymptote was presented. 
Starting from Feshbach molecules, high\hyp{}resolution one\hyp{}photon loss 
spectroscopy of the excited spin\hyp{}orbit coupled potentials revealed 26 
vibrational levels. The combination with published data from photoassociation 
spectroscopy led to the complete characterization of the long\hyp{}range 
part by near\hyp{}dissociation expansion expressions and improved $C_6$ 
coefficients. In a next step the data were combined with existing 
mass\hyp{}scaled data covering the short range of the $\textrm{B}^1\Pi$ 
potential. This allowed for the situation of empirical data covering the 
complete range of a molecular potential, which to our knowledge is unique for 
bi-alkali molecules. We additionally determined an updated value for the depth 
of the ${\textrm{B}^1\Pi}$ based on new spectroscopic data of the ground state 
potential. Hence, a complete empirical RKR potential was computed. 

The states of the spin\hyp{}orbit coupled $\Omega$=1$^{\textrm{up}}$ potential 
were investigated in particular, since it is directly connected to the 
$\textrm{B}^1\Pi$ in the short\hyp{}range. The experimentally resolved Zeeman 
splitting of the $v=-11$ vibrational sub\hyp{}level was used to identify the 
states of the $\Omega$=1$^{\textrm{up}}$ potential and the experimentally 
obtained g\hyp{}factor was found in good agreement with our theoretical 
prediction for Hund's case (c). We believe that the characterization of the 
$\Omega$=1$^{\textrm{up}}$ states is of particular importance for the purpose 
of finding a spectroscopic pathway for the transfer of the Feshbach molecules 
to the absolute ground state and the creation of a dipolar quantum gas. This is 
as at the inner turning point, the shallow $\textrm{B}^1\Pi$ potential offers 
excellent Franck\hyp{}Condon overlap with the ground state wave function at 
accessible wavelengths. As no hyperfine\hyp{}structure was resolved for the 
$\Omega$=1$^{\textrm{up}}$ state in ${}^{6}$Li${}^{40}$K, a similar approach as 
discussed in \cite{Yang2020} based on addressing a Zeeman component of the 
excited state by polarized light can be employed to control the hyperfine 
states for this pathway.

%\section*{Author Contributions}
%We strongly encourage authors to include author contributions and recommend 
%using \href{https://casrai.org/credit/}{CRediT} for standardised contribution 
%descriptions. Please refer to our general 
%\href{https://www.rsc.org/journals-books-databases/journal-authors-reviewers/author-responsibilities/}{author
% guidelines} for more information about authorship.

\section*{Conflicts of interest}
There are no conflicts to declare.
%In accordance with our policy on 
%\href{https://www.rsc.org/journals-books-databases/journal-authors-reviewers/author-responsibilities/#code-of-conduct}{Conflicts
% of interest} please ensure that a conflicts of interest statement is included 
%in your manuscript here.  Please note that this statement is required for all 
%submitted manuscripts.  If no conflicts exist, please state that ``There are 
%no conflicts to declare''.
%

%TC:break Acknowledgments
\section*{Acknowledgements}
%The Acknowledgements come at the end of an article after Conflicts of interest and before the Notes and references.
This research is supported by the National Research Foundation, Prime Ministers 
Office, Singapore and the Ministry of Education, Singapore under the Research 
Centres of Excellence program. We further acknowledge funding by the Singapore 
Ministry of Education Academic Research Fund Tier 2 (grant MOE2015-T2-1-098).

%%%END OF MAIN TEXT%%%

\balance


%%%REFERENCES%%%
%TC:break Bibliography
\bibliography{LongRangeSpectroscopy} %You need to replace "rsc" on this line with the name of your .bib file
\bibliographystyle{rsc} %the RSC's .bst file

%%%%%%%%%%%%%%%%%%%%%%%%%%%%%%%%%%%%%%%%%%%%%%%%%%%%%%%%%%%%%%%%%%%%%%%%%%%%%%%%%%%%%%%%%%%%%%%%%%%%%%%%%%%%%%%%%%%%%%%%%%%%%%%

\thispagestyle{plain}
	
%%%PAGE SETUP - Please do not change any commands within this section%%%
\makeFNbottom
\makeatletter
\renewcommand\LARGE{\@setfontsize\LARGE{15pt}{17}}
\renewcommand\Large{\@setfontsize\Large{12pt}{14}}
\renewcommand\large{\@setfontsize\large{10pt}{12}}
\makeatother
	
\makeatletter 
\renewcommand\@biblabel[1]{#1}            
\renewcommand\@makefntext[1]% 
{\noindent\makebox[0pt][r]{\@thefnmark\,}#1}
\makeatother 
\renewcommand{\figurename}{\small{Fig.}~}
\sectionfont{\sffamily\Large}
\subsectionfont{\normalsize}
\subsubsectionfont{\bf}
\sectionfont{\bf}
\setstretch{1.125} %In particular, please do not alter this line.
\setlength{\skip\footins}{0.8cm}
\setlength{\footnotesep}{0.25cm}
\setlength{\jot}{10pt}
\titlespacing*{\section}{0pt}{4pt}{4pt}
\titlespacing*{\subsection}{0pt}{15pt}{1pt}
%%%END OF PAGE SETUP%%%	
	
%%%FIGURE SETUP - please do not change any commands within this section%%%
%\makeatletter 
%\newlength{\figrulesep} 
\setlength{\figrulesep}{0.5\textfloatsep} 
	
%\newcommand{\topfigrule}{\vspace*{-1pt}% 
%	\noindent{\color{cream}\rule[-\figrulesep]{\columnwidth}{1.5pt}} }
	
%\newcommand{\botfigrule}{\vspace*{-2pt}% 
%	\noindent{\color{cream}\rule[\figrulesep]{\columnwidth}{1.5pt}} }
	
%\newcommand{\dblfigrule}{\vspace*{-1pt}% 
%	\noindent{\color{cream}\rule[-\figrulesep]{\textwidth}{1.5pt}} }
	
%\makeatother
%%%END OF FIGURE SETUP%%%

%%%TITLE, AUTHORS AND ABSTRACT%%%	
\onecolumn

\date{}
\title{\textbf{Supplementary Information:}\\
\textbf{Empirical LiK excited state potentials: connecting short range 
and near dissociation expansions}}
\maketitle	
\vspace{-1cm}		
	\noindent \Large{Sofia Botsi,\textit{$^{a}$} Anbang Yang,\textit{$^{a}$} 
	Mark M. Lam,\textit{$^{a}$} Sambit B. Pal,\textit{$^{a}$} Sunil 
	Kumar,\textit{$^{a}$} Markus Debatin,\textit{$^{a}$} and Kai 
	Dieckmann$^{\ast}$\textit{$^{a,b}$}}\\ 
			
	\noindent \large{\textit{$^{a}$~Centre for Quantum Technologies (CQT), 3 
	Science Drive 2, Singapore 117543}\\
	\textit{$^{b}$~Department of Physics, National University of Singapore, 2 
	Science Drive 3, Singapore 117542}\\
	\textit{$\ast$~Corresponding author E-mail: phydk@nus.edu.sg}}
	
%%%END OF TITLE, AUTHORS AND ABSTRACT%%%

%%%FONT SETUP - please do not change any commands within this section
\renewcommand*\rmdefault{bch}\normalfont\upshape
\rmfamily
\section*{}
\vspace{0cm}	

\renewcommand\thesection{S1}
\section{Spin\hyp{}orbit coupled potentials}
In order to obtain information about the composition of the excited electronic 
states, spin\hyp{}orbit coupling is considered in a simple approach to support 
the qualitative statements in the main text. A quantitatively more accurate 
description by a coupled\hyp{}channel calculation is beyond the scope of this 
analysis. For our purposes, we diagonalize the Hamiltonian:
\begin{equation}
	H_{\textrm{eff.}}=H_{\textrm{SO}}+H_{\textrm{pot.}}(\textbf{R})
	\label{eqn:THamiltonian}\quad ,
\end{equation}
\noindent where the term
$H_{\textrm{SO}}=H^+_{\textrm{SO}}+H^-_{\textrm{SO}}=\frac{a_{\textrm{SO}}}{2\hbar^2}(\textbf{s$_1$}\pm\textbf{s$_2$})\cdot\textbf{l}$
describes the spin\hyp{}orbit coupling interaction between open shell electrons 
and their own orbital angular momentum. The spin\hyp{}orbit coupling constant 
$a_{\textrm{SO}}$ is assumed to be independent of the internuclear distance 
$\textbf{R}$ and the value of the $4$P state of potassium is used 
%\cite{Tiecke2010}
[1]. This is a fair approximation as our earlier 
\textit{ab-initio} calculations suggest that $a_{\textrm{SO}}$ is varying by 
approximately a factor of two throughout the range of the potential allowing 
for the occurrence of mixed states at all binding energies (Supplementary 
Material of 
%\cite{Yang2020})
[2]. $H_{\textrm{pot.}}(\textbf{R})$ represents the 
bare potential curves in the Hund's case (a) eigenbasis 
%\cite{Allouche}
[3]. For 
simplicity the Zeeman effect is not taken into account. 

Fig.~S1 shows the projections of the Hund's case (c) coupled potentials onto 
the bare state basis resulting from the diagonalization. For the calculation of 
the Hund's case (c) g-factor presented in Section~3 of the main text, the 
long\hyp{}range composition of the $\Omega$=1$^{\textrm{up}}$ is of interest. 
From the figure one can see that only $\Pi$ states are relevant and hence only 
$L=1$ needs to be considered in the calculation. Further, it is apparent that 
the states $\Omega$=1$^{\textrm{up}}$ of the dyad and $\Omega=0^-$ of the upper 
triad contain a significant singlet component in the form of $\ket{^1\Pi}$ and 
$\ket{^1\Sigma}$ respectively. Consequently, these states are promising 
candidates of intermediate states which can facilitate the two\hyp{}photon 
transfer to the singlet absolute ground state. 

\newpage

\begin{figure}[hb]
	\renewcommand{\thefigure}{S1}
	\centering
	\includegraphics[width=18cm]{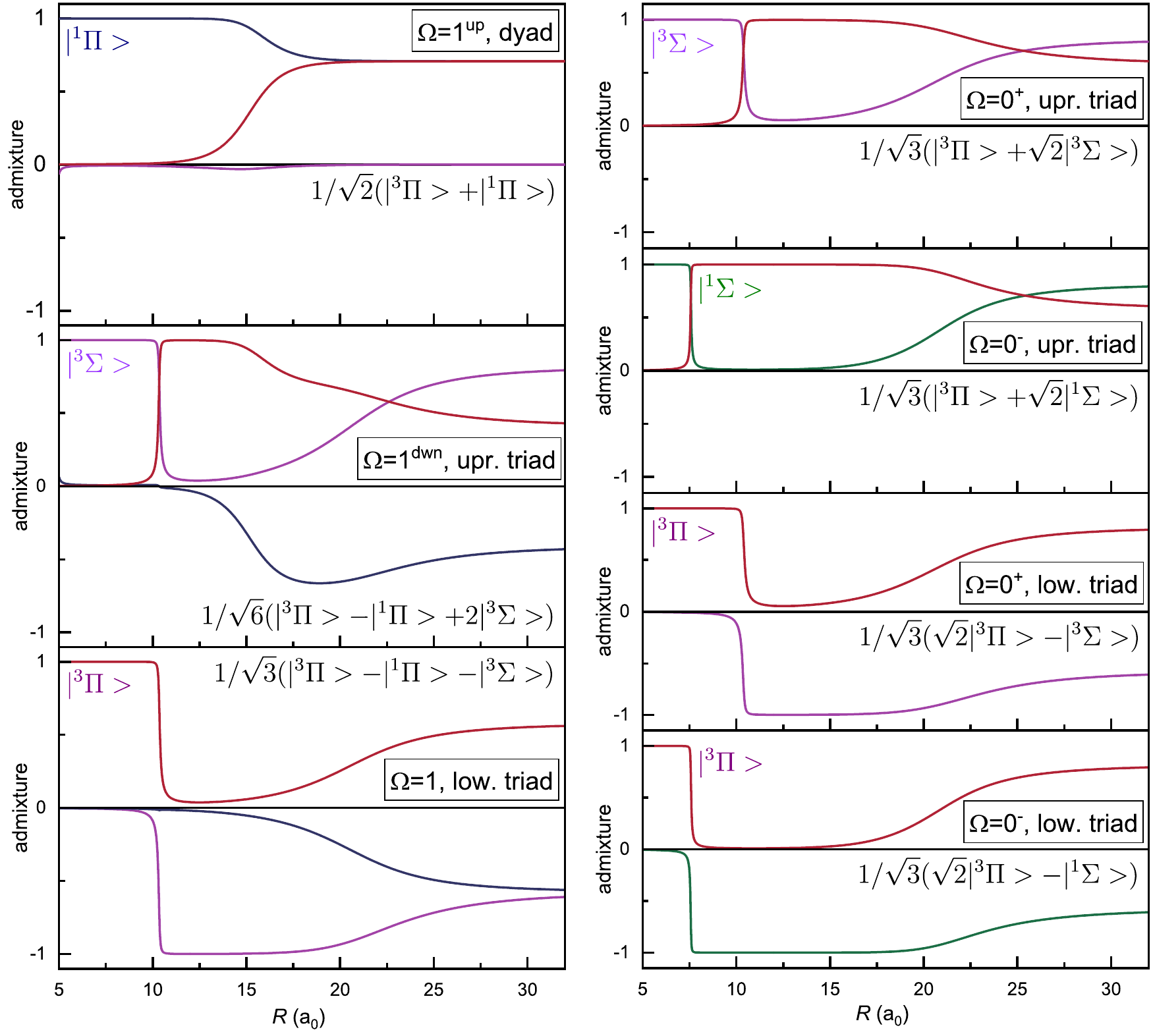}
	\caption{Projections of the spin\hyp{}orbit\hyp{}coupled Hund's case (c)
		potentials onto the bare Hund's case (a) states. The bare states are 
		labeled near the short\hyp{}range below the crossing points of about 
		$7.5\,\textrm{a}_0$ and $11\,\textrm{a}_0$. The long\hyp{}range state 
		composition is indicated by the formulas. For simplicity the full representations of 
		the symmetry of the states for the cases of $\Omega$ doubling as well as for the $\textrm{b}^{3}\Pi$ state are 
		omitted. The $\Omega=2$ state is not shown, since it does not couple}
	\label{fgr:S1}
\end{figure}

\newpage

\renewcommand\thesection{S2}
\section{Classical inner and outer turning points of the ${\textrm{B}^1\Pi}$ 
potential}

{ % begin box to localize effect of arraystretch change
	\renewcommand{\arraystretch}{1.5}
	\begin{table}[ht]
		\centering
		\renewcommand{\thetable}{S2}
		\small
		\caption{\ Fully empirical based RKR representation of the $\textrm{B}^{1}\Pi$ potential of the
		${}^{6}\textrm{Li}{}^{40}\textrm{K}$ molecule. The table contains the calculated classical inner and outer turning points with respective vibrational energies and rotational constants. Counting of the vibrational level index is downwards from the dissociation threshold}
		\label{tbl:RKR1results}
		\begin{tabular*}{0.48\textwidth}{@{\extracolsep{\fill}}lllll}
			\hline
			-v & $\textrm{R}_{\textrm{inner}}\,$(a$_0$) & 
			$\textrm{R}_{\textrm{outer}}\,$(a$_0$) & $G_v$ (cm$^{-1}$) & $B_v$ 
			(cm$^{-1}$)\\
			\hline
			1  & 5.5664 & 61.0125 & 1686.964 & 0.005\\
			2  & 5.5666 & 39.4987 & 1686.485 & 0.012\\
			3  & 5.5673 & 31.4806 & 1684.966 & 0.019\\
			4  & 5.5687 & 26.9993 & 1681.840 & 0.027\\
			5  & 5.5710 & 24.0573 & 1676.591 & 0.035\\
			6  & 5.5744 & 21.9495 & 1668.779 & 0.043\\
			7  & 5.5791 & 20.3545 & 1658.053 & 0.051\\
			8  & 5.5851 & 19.1013 & 1644.174 & 0.060\\
			9  & 5.5926 & 18.0886 & 1627.031 & 0.069\\
			10 & 5.6014 & 17.2496 & 1606.660 & 0.080\\
			11 & 5.6116 & 16.5351 & 1583.244 & 0.090\\
			12 & 5.6228 & 15.8998 & 1557.101 & 0.102\\
			13 & 5.6351 & 15.2831 & 1528.497 & 0.112\\
			14 & 5.6487 & 14.6576 & 1496.467 & 0.116\\
			15 & 5.6643 & 14.1715 & 1459.738 & 0.113\\
			16 & 5.6809 & 13.7386 & 1420.279 & 0.114\\
			17 & 5.6984 & 13.3072 & 1378.324 & 0.119\\
			18 & 5.7171 & 12.8860 & 1333.469 & 0.125\\
			19 & 5.7368 & 12.4776 & 1285.460 & 0.130\\
			20 & 5.7583 & 12.0822 & 1234.096 & 0.136\\
			21 & 5.7815 & 11.6983 & 1179.195 & 0.142\\
			22 & 5.8067 & 11.3240 & 1120.573 & 0.148\\
			23 & 5.8345 & 10.9567 & 1058.004 & 0.155\\
			24 & 5.8654 & 10.5940 &\ \ 991.168 & 0.162\\
			25 & 5.9001 & 10.2340 &\ \ 919.587 & 0.169\\
			26 & 5.9396 & \ \ 9.8763 &\ \ 842.579 & 0.176\\
			27 & 5.9849 & \ \ 9.5215 &\ \ 759.243 & 0.185\\
			28 & 6.0375 & \ \ 9.1715 &\ \ 668.490 & 0.193\\
			29 & 6.0997 & \ \ 8.8283 &\ \ 569.137 & 0.202\\
			30 & 6.1756 & \ \ 8.4932 &\ \ 460.092 & 0.211\\
			31 & 6.2733 & \ \ 8.1644 &\ \ 340.627 & 0.219\\
			32 & 6.4103 & \ \ 7.8315 &\ \ 210.781 & 0.227\\
			33 & 6.6436 & \ \ 7.4463 &\ \ \ \ 71.895& 0.232\\
			\multicolumn{3}{l}{\hspace{1.6cm}\textrm{R}$_\textrm{e}$=7.0169 a$_0$}\\
		\end{tabular*}
	\end{table}
} % end box
%%

%\newpage

%\bibliography{LongRangeSpectroscopy} %You need to replace "rsc" on this line 
%with the name of your .bib file
%\bibliographystyle{rsc} %the RSC's .bst file

\end{document}